# On the Feasibility of Social Network-based Pollution Sensing in ITSs

Rita Tse, Yubin Xiao, Giovanni Pau, Marco Roccetti, Serge Fdida and Gustavo Marfia

*Abstract*—Intense vehicular traffic is recognized as a global societal problem, with a multifaceted influence on the quality of life of a person. Intelligent Transportation Systems (ITS) can play an important role in combating such problem, decreasing pollution levels and, consequently, their negative effects. One of the goals of ITSs, in fact, is that of controlling traffic flows, measuring traffic states, providing vehicles with routes that globally pursue low pollution conditions. How such systems measure and enforce given traffic states has been at the center of multiple research efforts in the past few years. Although many different solutions have been proposed, very limited effort has been devoted to exploring the potential of social network analysis in such context. Social networks, in general, provide direct feedback from people and, as such, potentially very valuable information. A post that tells, for example, how a person feels about pollution at a given time in a given location, could be put to good use by an environment aware ITS aiming at minimizing contaminant emissions in residential areas. This work verifies the feasibility of using pollution related social network feeds into ITS operations. In particular, it concentrates on understanding how reliable such information is, producing an analysis that confronts over 1,500,000 posts and pollution data obtained from on-the-field sensors over a one-year span.

## I. Introduction

TRAFFIC is widely considered one of the main societal problems in many different countries of the world. Although high traffic levels are typically considered a sign of prosperity, and, as such, initially often welcomed, prolonged exposures to high traffic levels also bring in a number of economic and health issues that include: loss of productivity, drivers' stress, and a plethora of diseases caused by the excess of chemical agents that are released by combustion engines (e.g., microscopic particles, carbon monoxide, etc., [1], [2]). Clearly, many different countermeasures have been taken during the decades, which include, but are not limited to, implementing better public transportation systems [3], gradually substituting petroleum-based combustion engines with cleaner propellers (e.g., electric, ethanol or methane [4], [5]) and to the enforcement of congestion taxes [6] (for those who drive through given city areas during pre-defined time frames). In addition, Intelligent Transportation Systems (ITSs), systems capable of managing traffic resources (e.g., traffic light cycle times) and providing traffic related information to drivers (e.g., construction sites, traffic loaded roads) have flourished, with the aim of optimizing the use of available road and signaling infrastructures [7].

A wealth of research, in particular, has focused on the devise and implementation of ITSs, as their construction pose an important number of interesting problems in areas such as mathematical optimization [8], [9], [10], synchronous and asynchronous distributed communications [11], queuing theory [12], traffic congestion detection and forecasting [13], just to mention a few. Summarizing, ITSs have been conceived to measure traffic states in order to facilitate, through centralized (e.g., enforcing traffic light timings [14]) or decentralized means (e.g., providing traffic information to personal navigation systems [15]), traffic flows [16]. It is hence easy to understand why a fundamental role is played, in ITS, by traffic sampling systems [17] (i.e., those systems that measure and report the amount of traffic that is flowing through a given road), systems that exploit different type of technologies (e.g., cellular, sensor and vehicular ad hoc network data) to estimate traffic flows in real-time.

Now, with the progressive spread of a general environment awareness and the development of novel sustainable mobility paradigms, the reduction of pollution levels has moved from being considered one of the positive side effects produced by the use of ITSs to become one of their primary objectives [18], [19]. As a consequence, researchers have started investigating how pollution information could be used to modify and influence traffic control algorithms, giving environmental measurements a clear, policy-sensitive role in future traffic management and control schemes. Pollution sensors have, hence, become important feeds of information, just as important as traffic flow sensors, to be considered in the design of ITSs [20]. Such evolution should clearly push urban areas to deploy pervasive pollution sensor systems at large, as their output information could then be put to good use to improve traffic management and emission containment operations.

Although the cost of pollution sensing platforms is

R. Tse is supported by Macao Polytechnic Project N. (RP/ESAP-02/2014)-Bridging Urban Sensing and Social Networks. R. Tse and Y. Xiao are with the Macao Polytechnic Institute, Macao, China (e-mail: ritatse@ipm.edu.mo, p1207949@ipm.edu.mo). G. Pau is with the University Pierre et Marie Curie — Sorbonne Universities, Paris, France (e-mail: giovanni.pau@lip6.fr) / UCLA Computer Science Department. S. Fdida is with the University Pierre et Marie Curie Sorbonne Universities, Paris, France (e-mail: serge.fdida@lip6.fr). M. Roccetti and G. Marfia are with the University of Bologna, Italy (email: marco.roccetti@unibo.it, gustavo.marfia@unibo.it).

progressively decreasing, a pervasive and widespread deployment of such technologies is still far away in time for its costs. Both satellite and terrestrial solutions are being studied for a fine grained pollution estimation, with different results in terms of sensitivity, resolution and accuracy [22], [23], [24]. The common denominator, however, of all these technologies is the use of specialized hardware devices capable, by different means, of estimating the presence and concentration of specific pollutants. No system has instead involved so far any feedback received from the public. The reasons of such oversight may be mainly found in the following two points: (a) a rapid and large scale collection of information from the public is hard, and, (b) no well accepted proof regarding the reliability of such information is available (i.e., studies performed on limited groups of people revealed that actual pollution situations might be vey different from what people perceive [25]).

New lifestyles, habits and ways of communicating have, however, emerged, with the now extensive use of Online Social Networks (OSNs). With OSNs, personal communications are no more limited to one-to-one patterns of exchange of information (typical with phone calls and most emails), but often (i.e., in the case of public posts) follow a structure where information flows in a one-to-all fashion, allowing posts and comments to be read, commented and reposted by a multitude of users. In addition, posts can in principle touch upon any topics, as users can communicate anything with no censorship. All this to formulate the following conjecture: the big data of OSN user posts, that typically expose what a person thinks and how s/he feels and behaves, could be a very interesting source of pollution related information.

As a conjecture, this needs to be proven. In particular, it requires assessing whether and how frequently people touch upon pollution topics. In addition, even if a sufficient number of posts were available, their value would also need to be checked: could the observation of a given number of posts complaining about air quality really represent a red flag indicating that pollution conditions are not contained within their required limits? The quantity and quality of user posts need, hence, to be evaluated in order to design ITSs that may effectively exploit such information.

Explaining the contribution of this paper is, at this point, very simple, as it mainly aims at understanding: (a) how OSN data could be integrated into ITS operations, (b) utilizing any pollution related posts that may be found, (b) in case such posts effectively followed a pattern reflecting pollution conditions. To do so, we: (a) propose a novel ITS architecture that integrates OSN data, and, (b) verify the feasibility of utilizing user posts for traffic management purposes analyzing over 1,500,000 Sina Weibo (a very popular social networking platform in China) posts uploaded during an observation period of one year in two Chinese cities, Hong Kong and Guangzhou.

This paper is organized as follows. After reviewing the approaches that fall closest to the one that has been presented in Section II, how a traditional ITS architecture should be modified to accommodate novel information samples coming from social networks in Section III. We then move on to study how Sina Weibo Sina Weibo data relates to pollution sensor data in Sections IV and V. We finally conclude with Section VI.

## II. RELATED WORK

A wealth of research has been carried out on ITSs and their components beginning over three decades ago [26]-[35]. However, very little has considered the opportunity of putting to good use the information that could be found on social networks. The following briefly analyzes the few most representative examples.

The authors of VoiceTweet devised a system that leverages on Twitter feeds to communicate traffic information perceived by a driver [36]. In practice, a driver, when stuck in traffic, can share his/her driving experience recording a voice tweet with his/her social navigator. The social navigator then: (a) tags the message with its timestamp and the vehicle's location, and, (b) sends the message to a server that, in turn, groups together all the messages that are received from nearby locations during a given time frame. The server then periodically sends out tweet digests on social channels: when receiving such messages, social navigators prune off any unnecessary information (e.g., outdated or off route information) and compute a new route, or update the existing one. While VoiceTweet represents an innovation in the ITS field, its use requires the creation of a platform and of a wide user base. Such solution, hence, only marginally exploits the potential of OSNs, as these are mainly employed as a communication channel conveying the information created by a new service customer set, rather than an already available source of data.

Social networking has also been exploited, within the ITS context, to alleviate traffic through the implementation of a dynamic ride sharing community service. In practice, a recent work proposes the use of an information grid system, which, leveraging on OSNs, implements a carpooling service [37]. Pedestrians and drivers can utilize the proposed system uploading their schedule and locations of interest (i.e., origin and destination points) on OSNs. The system downloads such information from OSNs and seeks for the best time/location correspondences between pedestrians and drivers. If feasible solutions are found, both parties are informed and conveniently left with the freedom of taking advantage of the proposed opportunity, or not. This work, just as the previous one, exploits OSNs as communication means, rather than as independent sources of information.

Interestingly, both [38] and [39] discuss the opportunities that could be created with the integration of cooperative technologies and ITSs. In particular, Miller envisions a *virtual environment where data streams are fused, interpreted and made available with tools for human engagement and shared decision-making* [38]. Following a similar line of reasoning, Mahmassani acknowledges that ITSs have incredibly progressed with the advent of mobile platforms and apps, but also recognizes that there is a long way to go to take full advantage of the personalization/customization/socialization opportunities that they pose ahead of ITS planners and designers. While both works provide important intuitions, neither one nor the other practically explain how such ideas could be put to good use to improve existing ITSs. This work moves a first step in filling such gap.

### III. A Pollution Aware ITS Architecture

The proposal of a novel ITS architecture that may successfully integrate an OSN source of information requires an understanding of how such solutions work and of how they may evolve. For this reason the following subsection first provides a brief review of the state of the art and then proceeds identifying where such type of sources could be intermingled with others. Such approach is not sufficient, however, as the information collection process particularly depends on the patterns that emerge (if any) between the two groups of variables that are at stake, in particular: pollution-related posts and pollutant levels.

#### A. OSN Pollution Information Integration

Before describing which type of ITS architecture could effectively incorporate OSN-based sources of information, it is worthwhile summarizing how such complex systems are organized. Two main approaches have emerged with the aim of limiting the business and societal costs of vehicular congestion [35], [40], [41]. The first amounts to provide aggregate traffic information (e.g., intensity of traffic volumes estimated and lane occupancy rates estimated with video cameras and induction loops) to Advanced Traffic Management Systems (ATMSs) to control the road infrastructure (e.g., traffic light cycles or congestion charges) and to provide aggregate traffic information to drivers (e.g., with dynamic message signs and FM radios). The second approach is, instead, based on the idea of feeding road traversal times, sampled with probe vehicles, to Advanced Traveler's Information Systems (ATISs) which, in turn, supply single drivers with a feedback on traffic and with suggestions on the best routes (i.e., a driver's optimal route) to their destinations.

Both of such approaches deal with a plethora of traffic information sources (e.g., cellular networks, vehicular ad hoc networks, induction loops, video cameras, etc.) that have or are in the process of becoming available [42]. The idea is to also integrate data derived from OSNs, just as sensor data, into such infrastructure (Figure 1). In order to do this, for the particular case of pollution data, it is important to remind that the two systems are operated by players with completely different points of view and needs: transportation authorities run ATMSs, while ATISs are marketed by private businesses. The authorities which monitor the containment of pollution emissions could eventually require ATMSs to, for example, take online decisions depending on past, current or expected situations, discouraging traffic flows from traversing areas that were, are or are expected to be experiencing high pollution levels. ATISs could also be involved into such process (e.g., provide concerned drivers with sustainable routes, implement carpooling systems, indicate multi-modal transportation solutions that may jointly optimize travel time and sustainability-related variables) although with potentially many more difficulties, as the ultimate goal of any driver is that of diminishing the time s/he spends on the road.

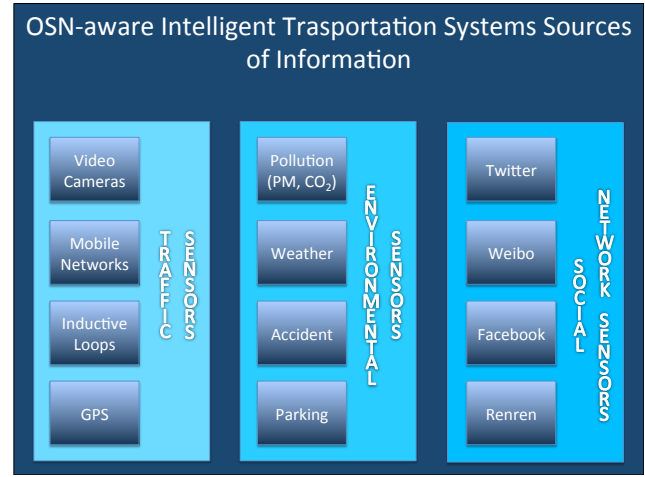

Fig. 1. A novel ITS information infrastructure integrating OSN sources.

#### B. OSN Posts and Traffic Pollutants Interconnections

Finding patterns that bind pollution-related OSN posts to chemical agent levels requires providing answers to two questions: (a) which are the pollution emissions and levels (if any) that push OSN users to write given types of posts, and, (b) if such posts exists, do they strictly touch upon pollution, or do they include, instead, references to other topics (e.g., weather, congestion, ability to breath, etc.)? An answer to both of such questions can be given proceeding in the following order: the candidate sentinel topics and the pollutants of interest are identified in this subsection, whereas the search for any existing patterns between such quantities is postponed to the experimental section of this work.

Now, the peculiarity of an OSN source (i.e., where the vast majority of information is provided by human beings), when compared to hardware sensors, is that a certain amount of interpretation is required. For example, a person complaining about haze (i.e., haze in general may be due to humidity and have nothing to do with pollution) could in reality be experiencing a highly polluted situation. The same may be said for other variables other than pollution, weather-related and not. It is, hence, important to individuate a set of pollution-related areas that may provide reliable pollution-sentinel variables. A non-exhaustive list of such areas is given by the following: (a) traffic situation (e.g., congestion, parking, accidents, etc.), (b) weather, and, (c) health, just to mention the most evident ones.

Traffic is widely recognized as the main causes of air pollution in urban areas, pollution that in some areas may become so disruptive to provoke heavy modifications to normal weather conditions (e.g., haze, fog, etc.) [43]-[46]. For this reason posts indicating high congestion levels (or also the occurrence of accidents which will eventually lead to congestion) and the presence of haze and foggy conditions could be, in reality, respectively indicating that pollution levels will be soon climbing or have already exceeded the limits. While it may be reasonable to consider posts complaining about traffic congestion or particular weather conditions, for example, as indicators of high pollution levels, the use of health related posts requires particular care. Specific health problems (e.g., asthma, etc.) are not taken here into account, as the subject is so critical and multifaceted (i.e., understanding the causal relationship at the root of given health impairments can require years, while ITS decisions are taken within seconds to at most hours from an event) that it deserves a stream of research on its own. In this particular area, hence, this work will limit to consider only those posts that explicitly signal the experiencing of respiratory problems due to poor air conditions (which are still related to the health of a person, but on a much shorter time scale).

Obviously, OSN variables are not the only ones at stake, pollutants are the true variables under observation. Although many different elements are classified as pollutants, as possible causes of diseases and physical impairments, the ones that may be directly related to vehicular traffic include chemical agents such as [47]:

- Sulfur Dioxide ($SO_2$), characterized by an irritating odor, combined with other elements can in the air can contribute to the production of haze and reduced visibility;
- Ozone ($O_3$), bluish color, in high concentrations its smell is sharp, resembling the smell of electrical equipment;
- Carbon Monoxide (CO), colorless, odorless and non-toxic;
- Nitrogen Oxides ($NO_x$), which include Nitric Oxide (NO), a colorless, odorless and non-toxic gas, and Nitrogen Dioxide ($NO_2$), a reddish-brownish gas with a pungent odor, an important component of city smog;

and particulate matter, which can also contribute to haze in urban contexts, such as:

- Respirable Suspended Particles ($PM_{10}$), particulate matter with a diameter of 10 $\mu m$ or less;
- Fine Suspended Particles ($PM_{2.5}$), particulate matter with a diameter of 2.5 $\mu m$ or less.

Now, although important connections exist between the listed pollutants and what a person could feel (e.g., pungent odor, hazy weather, etc.), finding coherent relations between posts that are centered on traffic, weather conditions, health problems and pollution topics and the presence of pollutants amounts to a task with important challenges. The most important one, which is intrinsically present in any system that relies on OSN sources of information, is: to what extent is it possible to trust what people say? People may be reporting incorrect impressions, not necessarily with irony or malice, but simply because individuals intrinsically perceive the same phenomena in different ways. This is not surprising: some people, for example, suffer most from high temperatures, others from low, and, other people simply never bother to complaint about any disturbing conditions they may be facing. So, if someone posts: "it's too hot!" is it possible to trust that warm weather conditions will be found where such exclamation was uttered?

IV. POLLUTION-OSN POST SOURCES OF DATA

Searching for relations between pollution levels and OSN posts requires, as a first step, finding reliable sources of information, which may be summarized as: (a) a large-scale source of geo-located posts, and, (b) pollution sensor data. Although necessary for a successful outcome, finding such sources is not sufficient as, for example, it could be possible that no pollution-related posts were ever published, showing a general disinterest for the subject. Even in the favorable case where pollution-related posts were instead available, along with the data produced by a pollution sensor, posts might still be published too far from the sensor, making any attempt to relate the behavior of the two sources unrealistic. Last, but not least, even if both pollution-related posts and sensor data were available for the same areas and in the same time frames, finding an algorithm that may automatically and reliably select pollution-related posts from the enormous corpus that is daily published in any of the most popular OSNs is not trivial.

*A. Selecting the sources*

For the reasons that have been briefly introduced in this Section, the selection process of the sources of data has entailed finding: (a) an OSN with an abundant number of posts published nearby accessible pollution sensors, in, (b) at least two different cities notable for their pollution emissions (i.e., with, possibly, appreciable differences in terms of average pollution levels). For the purposes of this study, both of such conditions have been met by the cities of Hong Kong and Guangzhou, in China. Hong Kong and Guangzhou are both very populated cities (7.2 and 8.5 million of inhabitants, respectively), and, are both cities where pollution levels can reach very high values. However, due to a number of factors (i.e., distance from the ocean, etc.), Hong Kong exhibits average pollution levels that are lower than Guangzhou [48].

In addition, both Hong Kong and Guangzhou are among the cities where registered users of China's most popular microblogging system, aka Sina Weibo, are most active [49]. Sina Weibo, a hybrid of Twitter and Facebook, features a penetration rate above the 30% of Chinese Internet users. In December 2012, Sina Weibo had 503 million registered users, and about 100 million messages were posted each day at that time. Most interestingly, more than 70% of Sina Weibo users daily utilize such service from mobile, thus, not only sharing their posts but also their positions. This makes Sina Weibo an ideal source, as in most cases, for the two cities of interest, it is possible to verify whether a posts that has been shared has been written anywhere nearby one of the pollution sensors taken here into consideration.

Another interesting fact that led to the choice of Hong Kong and Guangzhou is that they are characterized by multiple sources of pollution information. In particular, in Hong Kong, it is possible to access the hourly averaged information provided by the government's $SO_2$, $CO$, $O_3$, $NO_x$, $NO_2$, $PM_{10}$ and $PM_{2.5}$ sensors from multiple different sites. For the purpose of this work, the Central, Causeway Bay and Mongkok sites have been chosen, as they cover the downtown areas of the city. In Guangzhou, instead, a downtown sensor located inside and operated by the American Embassy provides $PM_{2.5}$ hourly information.

*B. Selecting the posts*

Starting in October 3$^{rd}$ 2012, through April 3$^{rd}$ 2014, approximately 640 thousand Sina Weibo posts written by 228,684 users and 910 thousand posts published by 505,033, have been recorded, respectively, in the areas of Hong Kong and Guangzhou. Of all this corpus of data, only those posts that have been written within a radius of 5 km from any of the accessible pollution sensors (i.e., a value chosen as a tradeoff between finding a sufficient number of posts and observing pollution conditions consistent with the reference sensors) have been considered. Now the final step required to be able to support a comparison between pollution levels and pollution-related posts was the selection of the latter. Pollution-related post selection has been conducted according to the strategy indicated in Section III.B, i.e., searching for any posts that were related to air pollution in any of the following four categories: (a) pollution, (b) weather, (c) traffic, or, (d) health. To do so in an automatic way, three different paths have been followed.

The first path that has been followed has been that of utilizing a well-accepted methodology in text classification, which is a Naïve Bayesian classifier [50]. This has been performed resorting to a popular Naïve Bayes implementation, namely PyMining, which supported the construction of a database of relevant features concerning the pollution, weather, traffic and health [51]. In order to test the soundness of the classifications operated with this algorithm, a total number of 1000 posts, randomly taken from those that have been classified as related to pollution, have been divided into 10 subsets of 100 posts each and checked by 10 independent human classifiers, respectively. Unfortunately, this check revealed that the average performance of this method was as low as 61%.

The second path has, hence, involved utilized a second popular classification mechanism, namely a Support Vector Machine (SVM) [52]. In order to classify the Chinese text, this strategy required walking through two successive steps, training and testing. During the training part, a training document containing relevant terms related to pollution was manually created and its words were segmented to formulate a model from the computed eigenvalues and term-frequency inverse-document-frequency matrix. The application of such model to the Sina Weibo posts led to an improved performance, compared to the use of Naïve Bayes. In fact, the same test that was utilized before for Naïve Bayes, in this case revealed an average performance of 88% (i.e., 88% of the posts classified as pollution-related where effectively pollution-related).

TABLE I
POLLUTION-RELATED TERMS DICTIONARY.

| Category | Terms (English translation) |
|---|---|
| Pollution | Pollution, poor air, gray sky, air quality, |
| Weather | Haze, fog, gray sky, bad weather |
| Traffic | Traffic jams, congestion |
| Health | Unable to breath |

The third and final path entailed utilizing a dictionary constructed utilizing the keywords found most frequently in pollution-related posts. Table I summarizes the keywords that have been used to individuate posts falling into each of the four categories. Interestingly, the use of such terms provided an average performance of 91%, hence higher than any of the machine learning algorithms employed so far. Due to this fact, the pollution-related posts that are utilized in the rest of this work have been initially selected utilizing this last methodology. As a further and final check, however, additional tests have been performed on the final sets of data (ca. 15,000 posts in Hong Kong and almost 4,000 posts in Guangzhou), in order to ensure that the obtained posts were effectively pollution-related and avoid jeopardizing the rest of this work. Figures 2 and 3 provide a visual representation of where the pollution sensors are and where the majority of pollution-related posts appeared in Hong Kong and Guangzhou, respectively. Now, the experiments described in the next Section assess whether any relation existed between the pollution-related posts that were published during the period of observation and the pollution levels that were lived during that same time span in the areas of interest.

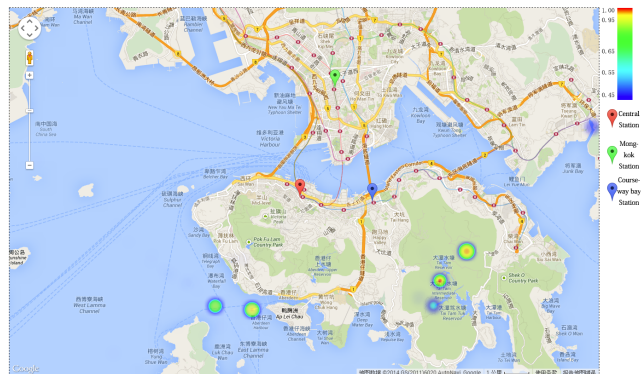

Fig. 2. Pollution-related posts and sensor positions in Hong Kong.

## V. FEASIBILITY EVALUATION

Evaluating the feasibility of detecting elevated pollution conditions from pollution-related posts requires finding any patterns connecting the two quantities. To do so, however, a hypothesis regarding how a user behaves after being exposed to elevated pollution levels is needed. The hypothesis that is considered in this work is: when exposed to high pollution, a person will, with a high chance, first of all try to find a way of getting out of such distressing situation. Once in a safe area (e.g., indoors), that person might then signal the occurrence of such situation.

Now, one of the problems that derive from such hypothesis is that the longer anyone waits to signal his/her distress due to experienced pollution conditions, the lower the value of such information for effective ITS operations. An ITS, in fact, needs to be able to take decisions based on information that is no more than a few hours old.

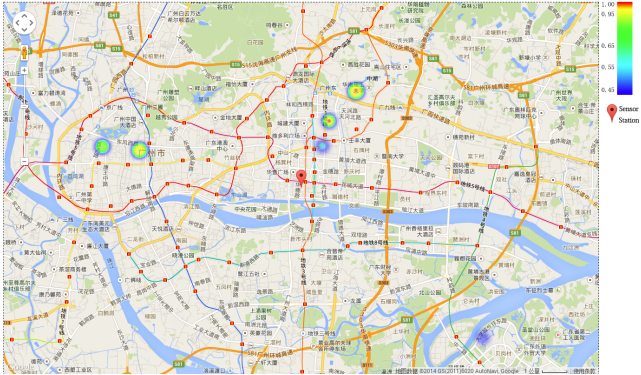
Fig. 3. Pollution-related posts and sensor positions in Guangzhou.

A tradeoff between the hypothesized user behavior and ITS requirements provides a methodology with which it is possible to seek for a correspondence between pollution-related posts and pollution sensed values: for each pollution-sensed value, count the number of pollution-related posts that have been published during the next T hours that follow. Considering that such information should be integrated into an ITS, the value of T has been set to 2 hours.

In the following hence, the results that are provided are how key pollution level statistical values vary, as the minimum number of posts published in the next 2 hours increases. Summarizing the results that have been obtained utilizing such methodology, it is possible to anticipate that an insufficient number of posts have been published in the areas of traffic and health and for this reason the remainder of this Section will concentrate on the most significant results obtained with the posts that contained pollution and weather key terms.

*A. Pollution keys*

When utilizing pollution keys, it has been possible to observe that the key statistics of certain pollution indictors varied as the minimum number of posts published increased. This, in particular, happened in downtown Hong Kong at the Central pollution station with Carbon Monoxide (as shown in Figure 4) and in downtown Guangzhou at the American Embassy with Fine Suspended Particulates (Figure 5), where minimum, first quartile, median, third quartile statistics and any outliers (i.e., empty dots) are plotted for a given minimum number of posts. When, at least 5 pollution posts have been published in Central Hong Kong, CO values remain confined between a minimum of almost 0.5 $mg/m^3$ and a maximum of 1.5 $mg/m^3$. When, instead, at least 5 pollution posts have been published in Guangzhou, PM$_{2.5}$ values remain confined between a minimum of 100 and a maximum close to 250 $\mu g/m^3$. The US Environmental Protection Agency (EPA) standard limits provides a better understanding of these values, which recommends the average concentration value of CO should not exceed the hourly average value of 40 $mg/m^3$ and the average concentration of PM$_{2.5}$ should not exceed 35 $\mu g/m^3$ over a 24 hours time frame (i.e., for the sake of completeness, such value is typically lower than the maximum value admitted over an hour averaging) [53]. Interestingly, the trends shown in Figures 4 and 5 reveal that CO values in Hong Kong are well below the EPA recommended limits, while the PM$_{2.5}$ are not.

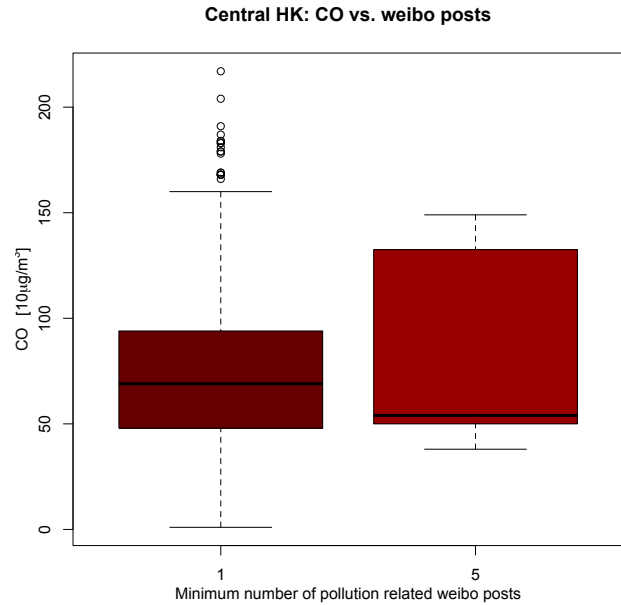
Fig. 4. Pollution posts vs. CO values in Central, Hong Kong.

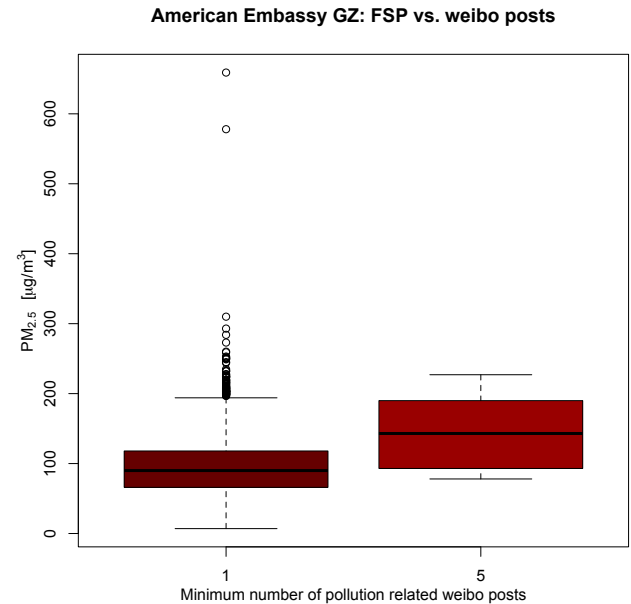
Fig. 5. Pollution posts vs. PM$_{2.5}$ values at the American Embassy, Guangzhou.

*B. Weather keys*

The great majority of posts that were mined utilizing weather-related keys let situations emerge where people complained of haze. Disturbing haze conditions, as anticipated in Section III.B, are consistent with the presence of high levels of given pollutant (e.g., NO$_2$, NO$_x$, SO$_2$ and particulate matter). In the following, the pollutants that exhibit an increase in any of their statistics as the minimum number of posts increases are shown.

As expected, as Nitrogen Dioxide (this same trend has been observed also with Nitrogen Oxides) is one of the main causes of haze in urban areas, its minimum value increases as the minimum number of posts increase (Figures 6, 7 and 8). However, considering that EPA recommends a $NO_2$ concentration limit equal to 100 ppb (188 $\mu g/m^3$) over an hour time frame, it is possible to observe that an increased number of posts do not substantially indicate that such value has been exceeded.

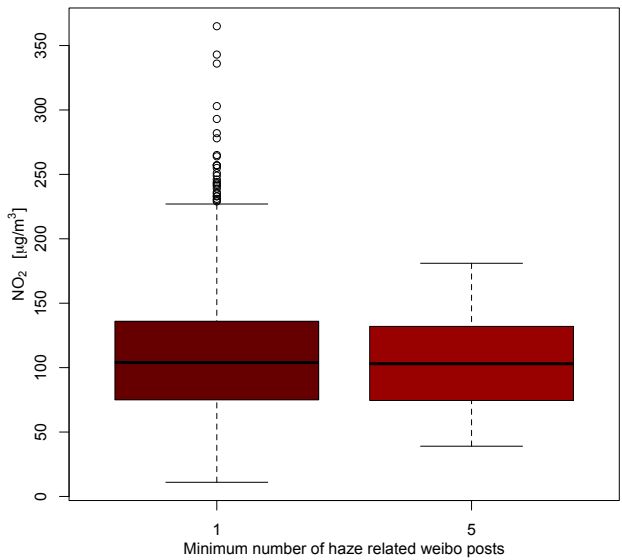

Fig. 6. Weather posts vs. $NO_2$ values in Causeway Bay, Hong Kong.

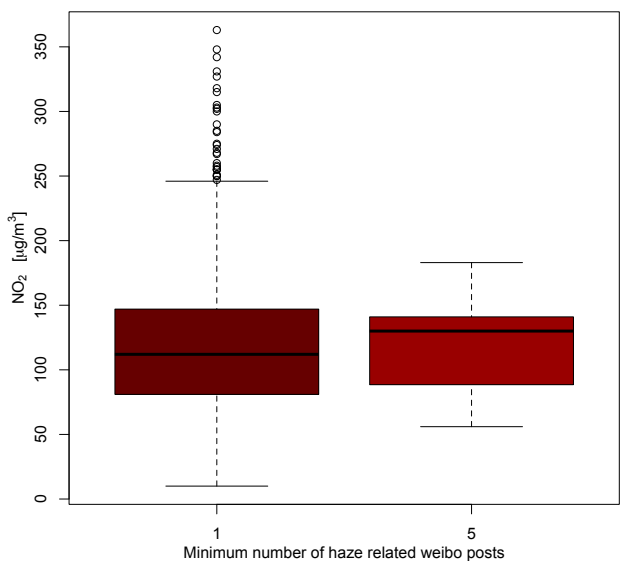

Fig. 7. Weather posts vs. $NO_2$ values in Central, Hong Kong.

It is, instead, very interesting to observe what happens with $PM_{2.5}$ data, when confronting the values obtained in Hong Kong and Guangzhou. In particular, Figures 9, 10 and 11 reveal no particular trend in the pollutant's statistics, as the number of posts increases. In all of the three cases, however, it is also possible to observe that the pollutant's values are more or less equally shared above and below the threshold limit of 35 $\mu g/m^3$ (EPA's limit for a 24 hour time frame), as median values fall in the 30-40 $\mu g/m^3$ range. As the number of post increase, however, it is possible to observe from the results obtained in Guangzhou (Figure 12) that a higher number of posts (i.e., at least 10 haze related posts) emerge as the median value of $PM_{2.5}$ falls way above 100 $\mu g/m^3$, hence, well beyond an acceptable value for health. A closer look at such data reveals that the days when Sina Weibo users in Guangzhou noticed most pollution levels, publishing at least 10 posts where they complained about haze, are the 13[th] of December 2013, the 8[th] and the 25[th] of January, the 17[th] 18[th] and 26[th] of February and the 3[rd] and 12[th] of March 2014.

As a final result, it has been possible to observe, again, a weak dependence between the values of CO concentration and the number of weather-related posts published by Sina Weibo users. In particular, Figures 13, 14 and 15 show this trend, which is more evident in Central, Hong Kong, than at other sites. Although weak, it is interesting to observe such trend, as unexpected, due to the fact that Carbon Monoxide is not one of the pollutants that can be accounted for haze.

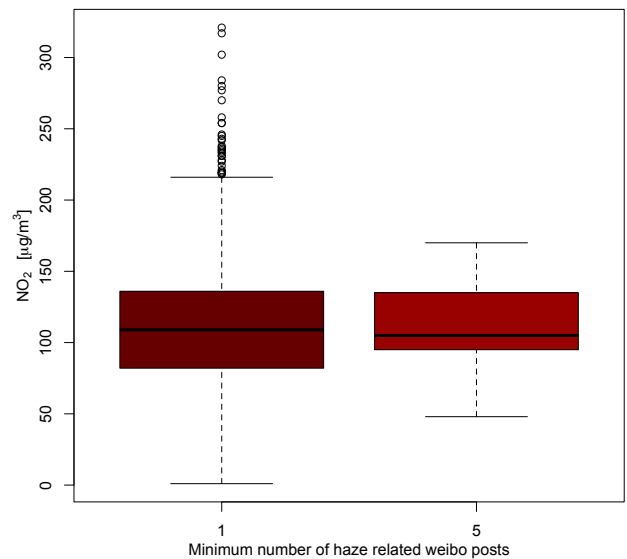

Fig. 8. Weather posts vs. $NO_2$ values in Mongkok, Hong Kong.

*C. Discussion*

From the results shown in this Section it emerges that a trend exists between the amount of Sina Weibo posts that are published, regarding given arguments, and the pollutant's levels recorded in adjacent areas. Unlike what one could possibly expect (i.e., the existence of an observable trend between posts complaining about traffic or pollution and pollutant recorded levels), the strongest bounds that have been found in this work are between pollution levels and haze. This is a key finding, as it means that, at least in the cities of Hong

Kong and Guangzhou in China, the concern for a particular weather condition like haze is high and that such concern can be associated with the observation of high pollution levels. In addition, what makes this finding particularly interesting, for the creation of an ITS solution that may rely on such type of technology, is the fact that the time frame within which the dynamics of posts occurs is 2 hours, hence a reasonable value within which practical traffic engineering pollution countermeasures could be taken.

Clearly, the results that are here provided also show the main shortcoming of this methodology: many situations where a high level of pollutants is recorded are not captured by an increase in the number of Sina Weibo posts (e.g., Figure 12 shows how the maximum value of $PM_{2.5}$, 659 $\mu g/m^3$, is observed within a 2-hour interval where only a single post was published). This is a shortcoming of this methodology, as a further observation of all graphs also shows how a pollutant's value can be extremely variable when only one relevant post was published: the fact that at some time a single post complaining about haze was published does not provide any reasonable guarantee that the pollution levels were of any interest or worry.

The strongest indication that this work provides, hence, in the field of OSN-based pollution information integration in ITS infrastructures is that such opportunity is solid and very interesting, as OSN contributors provide interesting feedback on what happens in their surroundings. This means that human beings, through the use of OSNs, could be effectively exploited as environmental sensors. Although a very interesting opportunity, this should however be accomplished considering the fact that human sensors come with a lot of randomness and noise and, as such, the pollution information that can be derived from the analysis of OSNs should be, from time to time, corroborated with other means of assessing pollution levels.

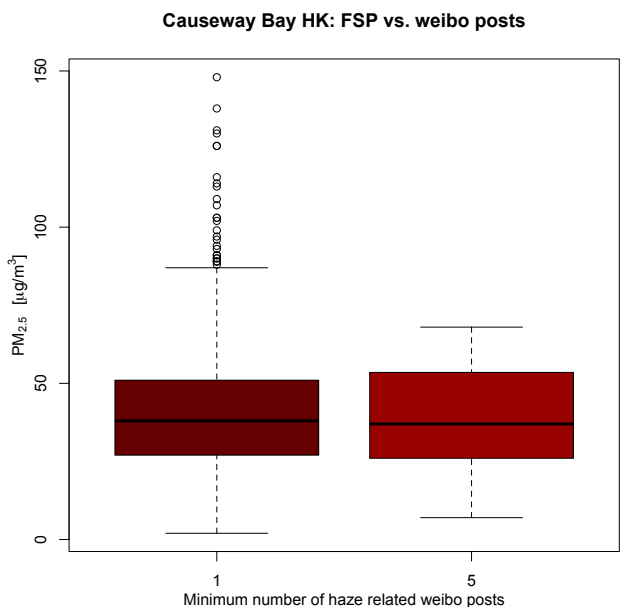

Fig. 9. Weather posts vs. $PM_{2.5}$ values in Causeway Bay, Hong Kong.

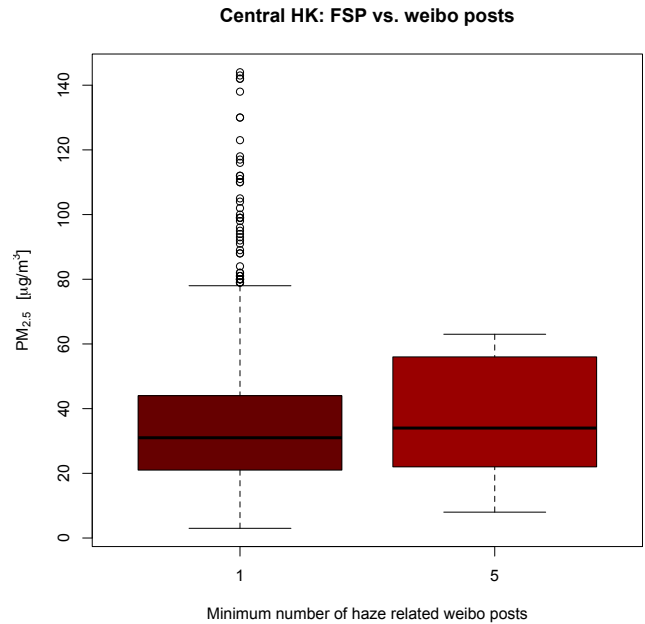

Fig. 10. Weather posts vs. $PM_{2.5}$ values in Central, Hong Kong.

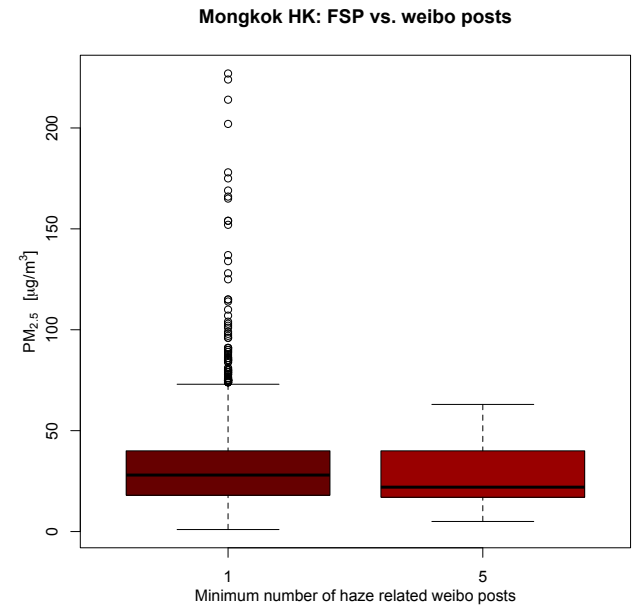

Fig. 11. Weather posts vs. $PM_{2.5}$ values in Mongkok, Hong Kong.

## VI. CONCLUSION

This is the first work that attempts to include an Online Social Network based source of information into an Intelligent Transportation System infrastructure. As such, the step that has been performed in this study is that of verifying the feasibility of using humans as sensors [54], applying such technique to the detection of pollution. In fact, a successful integration can only occur then moment that the public effectively utilizes OSNs the presence of pollution. The results that are here shown are promising, as a trend can be observed between the number of pollution-related posts that are published and the observed emission levels.

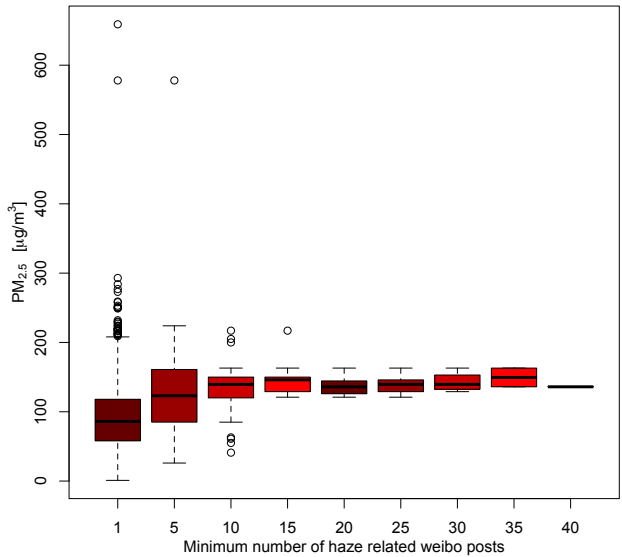

Fig. 12. Weather posts vs. PM$_{2.5}$ values at the American Embassy, Guangzhou.

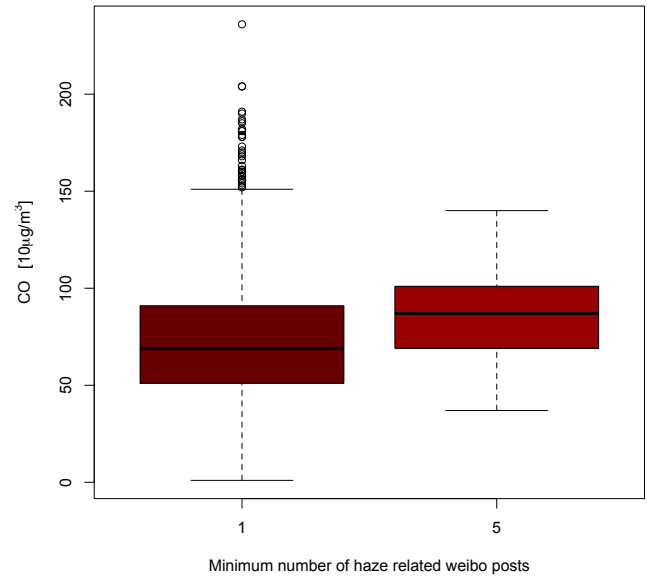

Fig. 14. Weather posts vs. CO values in Central, Hong Kong.

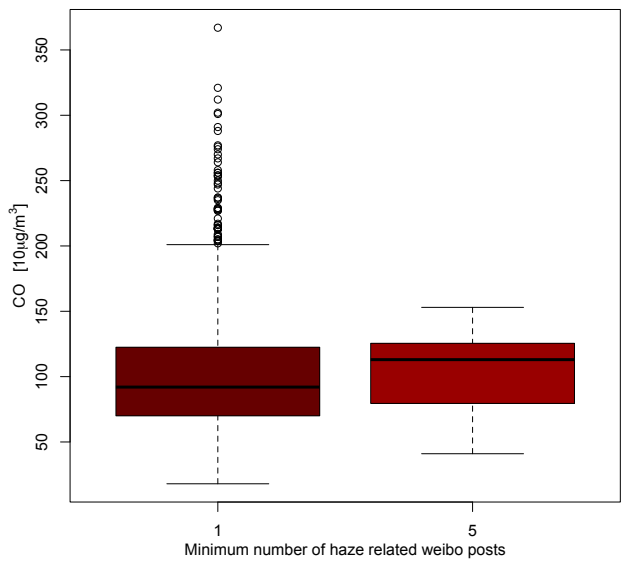

Fig. 13. Weather posts vs. CO values in Causeway Bay, Hong Kong.

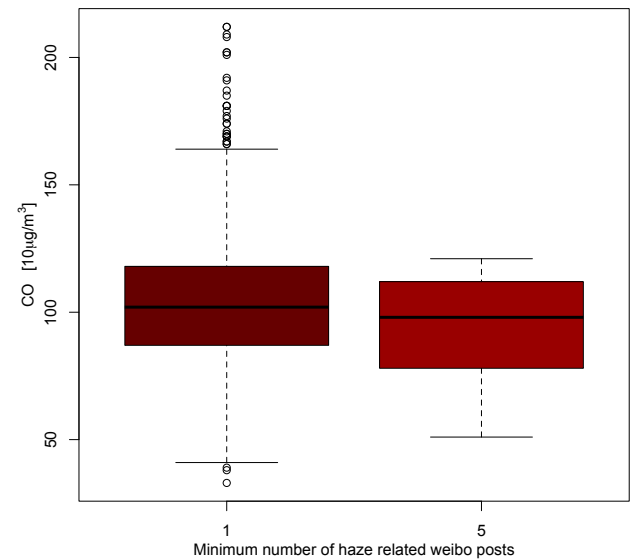

Fig. 15. Weather posts vs. CO values in Mongkok, Hong Kong.